\documentclass[showpacs, preprintnumbers, amsmath, amssymb,twocolumn]{revtex4}

\usepackage{graphicx}
\usepackage{dcolumn}
\usepackage{bm}
\usepackage{subfigure}

\begin{document}

\preprint{...}

\title{Convective Boson-Fermion pairing model constructed by oscillating one-dimensional optical superlattice}

\author{Tieyan Si}
\affiliation{ Physcis department, School of Sciences, and Key Laboratory of Microsystems and Microstructures Manufacturing-Ministry of Education,
Harbin Institute of Technology, Harbin, 150080, China }

\date{\today}

\begin{abstract}

Boson-fermion mixture exist in nature as quark-gluon plasma and $^3$He-$^4$He mixture. We proposed a convective boson-fermion pairing theory, that can be implemented by ultracold atoms in optical superlattice transformation between different configurations. This transformation may induce the collision and division between boson and fermion, which defines a theoretical convective pairing state. The paring Hamiltonian is Hermitian but it always generate a complex energy spectrum. Each finite gap state can be classified by a topological winding number. The stable pairing state only exists for certain discrete momentum vector zones. An unstable linear dispersion connects two neighboring stable pairing states. The boson-fermion gap function controls the momentum gap space between two neighboring pairing state. The critical temperature of transition from a gapped to gapless phase shows a maximal value at negative fermion chemical potential. The density of state for the pairing excitation diverges at low energy, thus most pairing states are observable at low energy.

\end{abstract}


\maketitle

\section{Introduction}

Optical lattice, generated by the interference of counter-propagating laser beams, provides an effective experimental technology to create many artificial quantum states, like superfluid, Mott-insulator, supersolid, quantum Hall state, and so on \cite{Bloch}. Boson-Fermion mixture is one of many interesting quantum matter that can be realized by ultracold atoms, and attracted broad research interest\cite{patu}\cite{cui}\cite{Ufrecht}\cite{wu}\cite{Kharga}\cite{Dehkharghani}\cite{Ikemachi}\cite{Brackett}\cite{HHu}\cite{guo}\cite{Rzhang}\cite{zfxu}\cite{Barbut}. The Boson-Fermion mixture that exist in nature is quark gluon plasma, $^3$He-$^4$He mixture, or $^{6}$Li-$^{7}$Li. It was theoretically suggested that two of the three quarks inside a proton may form color Cooper pair. The composite boson of color Cooper pair and the rest quark maybe can exist as boson-fermion mixture \cite{Rajagopal}. The theoretical possibility of a transition from Fermi gas of three quarks to boson-fermion mixture of quarks was revealed by T-matrix method \cite{Storozhenko}.

In recent years, boson-fermion mixture of ultracold atom is already implemented in laboratory by ultracold atom in optical lattice which traps $^{40}$K and $^{41}$K atoms gas \cite{Gunter}, and outside scope of the optical lattice techniques \cite{Truscott}. Boson-fermion mixture models predict many interesting collective phenomena, like superfluidity \cite{Matsyshyn}, supersolid phase \cite{orth}, phase separation \cite{Buchler}, boson-fermion cloud collapse \cite{chui}, weak-coupling pairing mode of boson and spin-polarized fermion \cite{Barillier}, and the supersymmetric Goldstino mode \cite{yueyu}\cite{Blaizot}.

Here we propose a convective pairing theory of boson-fermion mixture that is likely implemented by ultracold atom gas trapped in a superlattice of asymmetric double well potential. As this optical superlattice transforms between different potential configurations, the boson fermion would be collide in one trapping well or divide into two potential wells. This artificial boson fermion pairing state extended the wellknown Bardeen-Cooper-Schrieffer(BCS) pairing model in momentum space. However, unlike the BCS model, a chiral linear dispersion exist between neighboring energy loops of stable pairing state. In this paper, we focus on the topological characterization of the complex energy spectrum and critical temperature of phase transition from gapped phase to gapless phase.

\section{The boson-fermion pairing model}

\begin{figure}[htbp]
\centering
\par
\begin{center}
$
\includegraphics[width=0.37\textwidth]{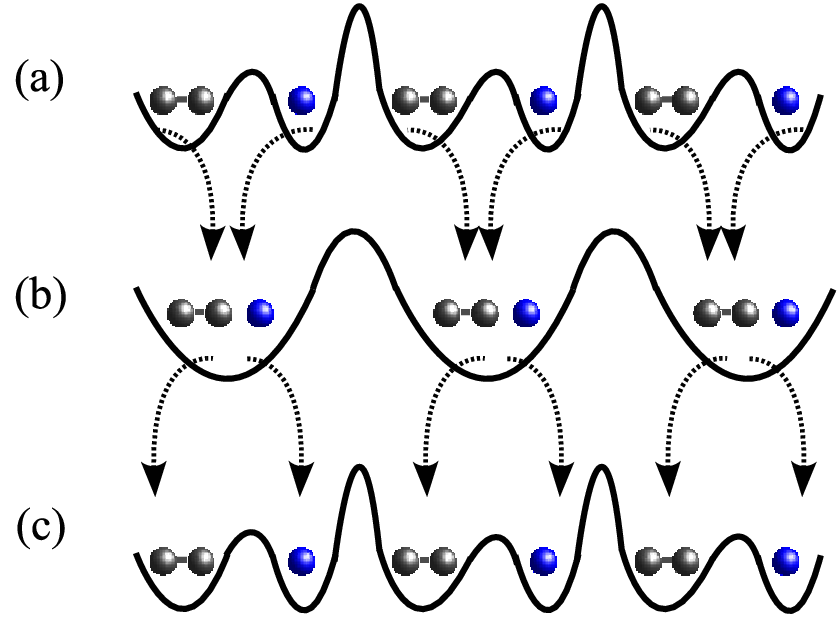}
$
\end{center}\vspace{-0.3cm}
\caption{\label{ring} (a) Odd number of atoms are grouped into a boson cluster (the dimer of two black balls) and fermion (the single blue balls) by the asymmetric double wells, which form a optical superlattice. (b) Transforming the superlattice of double well into a super lattice of single well potential drives the boson and fermion together to collide. (c) The inverse transformation from a single well to double well would separate the three atoms into one boson cluster and one fermion, and make them split into opposite directions.}
\vspace{-0.2cm}
\end{figure}

Ultracold-atoms in two dimensional optical superlattices has provided a well-developed technology to simulate d-wave Cooper pairs \cite{rey}. For a simpler case, one dimensional optical superlattice can also confine boson-fermion mixture and construct pairing state. The counter-propagating laser beams can be designed into different output patterns so that it generates a superlattice of asymmetric double wells or single wells (Fig. \ref{ring}). Initially one boson cluster and one fermion cluster are trapped in a super lattice of asymmetric double potential wells (Fig. \ref{ring}(a)). One approximation of single asymmetric double well is $U(x) = a x^4-b x^2+c x, $ (a=1, b=15, c=20). An optical superlattice is a tranlational copy of this double well over the whole space. When the asymmetric double potential well transforms into single harmonic potential well $U(x) = b x^2,$ the boson and fermion would collide each other (Fig. \ref{ring}(b)) to generate one big cluster with odd number of ultracold atoms. When the single potential well transformed back into the asymmetric double potential well, the big cluster splits into one fermion cluster and one boson cluster which are trapped in the two sub-wells separately (Fig. \ref{ring}(c)). A periodical transformation between these two types of optical superlattice pattern generates the boson-fermion pairing state, which is described by the effective Hamiltonian,
\begin{eqnarray}
&H&=\sum_{\langle{ij}\rangle}[t_{b}b^{\dag}_{i}b_{j}+t_{f}f^{\dag}_{i}f_{j}
-V(b^{\dag}_{i}f^{\dag}_{i}b_{i-e_{x}}f_{i+e_{x}}\nonumber\\
&+&b^{\dag}_{i-e_{x}}f^{\dag}_{i+e_{x}}b_{i}f_{i})+h.c.]-\sum_{i}[\mu_{b}b^{\dag}_{i}b_{i}+\mu_{f}f^{\dag}_{i}f_{i}].
\end{eqnarray}
Under Fourier transformation, $b_{i}=\frac{1}{\sqrt{N}}\sum_{k}e^{ikx_{i}}b_{k}$ and $f_{i}=\frac{1}{\sqrt{N}}\sum_{k}e^{ikx_{i}}f_{k}$, it yields the pairing Hamiltonian in momentum space,
\begin{eqnarray}
H=\sum_{k}[\epsilon_{k,b}b^{\dag}_{k}b_{k}+\epsilon_{k,f}f^{\dag}_{-k}f_{-k}]
-\sum_{k,k'}Vb^{\dag}_{k'}f^{\dag}_{-k'}b_{k}f_{-k}.\;\;
\end{eqnarray}
This pairing Hamiltonian share similar form as the well-known BCS model for superconductors, except here the pairing interaction only holds between boson and fermion. The kinetic energy of the boson and fermion reads
\begin{eqnarray}
\epsilon_{k,b}=-\mu_{b}-2t_{b}\cos(k_{x}),\;\;\;
\epsilon_{k,f}=-\mu_{f}-2t_{f}\cos(k_{x}).\;\;
\end{eqnarray}
The energy gap for exciting the boson-fermion pair is $\Delta_{bf}=\sum_{k'}V\langle{b_{k'}f_{-k'}}\rangle.$ The boson-fermion pairing model is quadratic in the expression of the pairing gap function,
\begin{eqnarray}
H&=&\sum_{k_{i}}[\epsilon_{k,b}b^{\dag}_{k}b_{k}+\epsilon_{k,f}f^{\dag}_{-k}f_{-k}]
+\Delta_{bf}^{\ast}\Delta_{bf}\;\;\;\;\nonumber\\
&-&\sum_{k}(\Delta_{bf}^{\ast}
b_{k}f_{-k}+\Delta_{bf}b^{\dag}_{k}f^{\dag}_{-k}).
\end{eqnarray}
This Hamiltonian describes a composite pair of a spinless boson and a spinless
fermion. The pair of boson and fermion is generated and annihilated
simultaneously. The boson always has an opposite momentum vector as
fermion. The energy gap is the order parameter in the framework of Gintzburg-Landau theory, the gap function ${\Delta}_{bf}$ is essentially the wave function of boson-fermion
pairs. In order to to meet the physical reality, ${\Delta}_{bf}$ should be an ordinary real function instead of Grassmann number. The total
particle number of boson-fermions pair is $\Delta^{\ast}_{bf}\Delta_{bf} = |\Delta_{bf}|^{2} = N_{bf}$. The total particle number of this boson-fermion pairing model is not conserved. Similar to the supersymmetric model of Boson-Fermion mixture \cite{yueyu},
we can also introduce a supersymmetry operator,
$Q=\sum_{k}b^{\dag}_{k}f_{k},$ $Q^{\dag}=\sum_{k}b_{k}f^{\dag}_{k}$.
The two supersymmetric operator obey the commutator,
$\{Q,Q^{\dag}\}=\sum_{k}(b^{\dag}_{k}b_{k}+f^{\dag}_{k}f_{k})=N_{total}.$
This supersymmetry operator does not commute with the
boson-fermion pairing Hamiltonian, $[Q,H]\neq0$. This model itself
does not keep supersymmetry. But the Hamiltonian still bears a hidden symmetry. If the boson and fermion have the same dispersion, $\epsilon_{k,b}=\epsilon_{k,f}$,
and in the meantime we perform a time reversal transformation on
momentum vector, $k\leftrightharpoons-k$, then the Hamiltonian would
keep the same formulation as before. This symmetry is time reversal symmetry jointed with supersymmetry. It will be shown in the following the boson-fermion pair remains stable and static in this jointed symmetry phase.

The eigenenergy of boson-fermion pair excitations is computed by Green-function method, which generates two energy branches,
\begin{eqnarray}\label{e1e2}
E_{1}&=&\left[\epsilon_{-}-\sqrt{[\epsilon_{+}]^2
-\Delta^{2}}\right],\nonumber\\
E_{2}&=&\left[\epsilon_{-}+\sqrt{[\epsilon_{+}]^2
-\Delta^{2}}\right].
\end{eqnarray}
where $\epsilon_{-}=\epsilon_{b}-\epsilon_{f}$ and $\epsilon_{+}=\epsilon_{f}+\epsilon_{b}.$ i. e.,
\begin{eqnarray}\label{E+E-}
\epsilon_{-}&=&\mu_{f}-\mu_{b}+(2t_{f}-2t_{b})\cos(k_{x}),\nonumber\\
\epsilon_{+}&=&-(\mu_{f}+\mu_{b})-(2t_{f}+2t_{b})\cos(k_{x}).
\end{eqnarray}
The energy gap for exciting up a pair of
boson-fermion depends on the occupation of fermion,
\begin{eqnarray}\label{gap}
\Delta^{2}=(2n_{f,-k}-1)(\Delta^{\ast}_{bf}\Delta_{bf}).
\end{eqnarray}
The density distribution of boson-fermion pair is $\Delta^{\ast}_{bf}\Delta_{bf}=\Delta^{2}=N_{bf}$. The gap is negative when the
occupation of fermions becomes zero, $n_{f,k}=0$, and is
positive if $n_{f,k}=1$. At half-filling state, $n_{f,k}=1/2$, it leads to the gapless phase. Since the particle number must be real number, a real physical state only exist above half-filling state, $n_{f,k}>1/2$. The energy difference between the two spectrum branches reads, ${\delta}E = E_{1}-E_{2}$,
\begin{eqnarray}\label{deltaE}
{\delta}E=\sqrt{[-(\mu_{f}+\mu_{b})-(2t_{f}+2t_{b})\cos(k_{x})]^2-\Delta^{2}}.
\end{eqnarray}
The energy difference ${\delta}E$ always has imaginary part unless the pairing gap energy is zero.

\begin{figure}[htbp]
\centering
\par
\begin{center}
$
\includegraphics[width=0.43\textwidth]{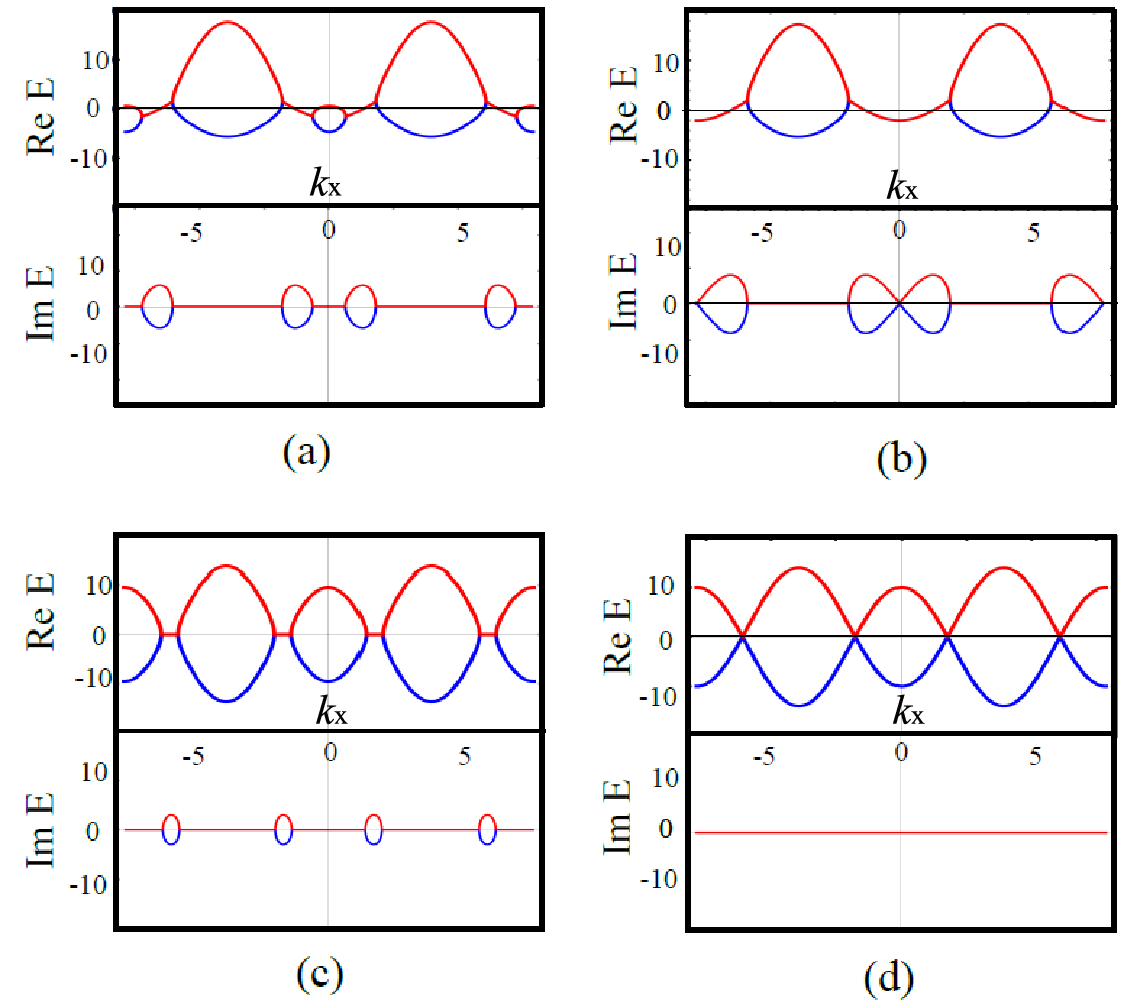}
$
\end{center}\vspace{-0.3cm}
\caption{\label{omega} The spectrum of the eigen-excitations of
with respect to different energy gaps (a) $\Delta = 3$, $\mu_{f}=-1,\mu_{b}=-3$, $t_{f}=1,t_{b}=3$; (b) $\Delta = 4$, $\mu_{f}=-1,\mu_{b}=-3$, $t_{f}=1,t_{b}=3$; (c) The spectrum for the supersymmetric case, $\Delta = 3$, $\mu_{f}=\mu_{b}=-1$, $t_{f}=t_{b}=3$; (d) $\Delta =0$. $\mu_{f}=\mu_{b}=-1$, $t_{f}=t_{b}=3$. Here the occupation of fermions is larger than 1/2.}
\vspace{-0.2cm}
\end{figure}

The eigenenergy spectrum of the boson-fermion pairing always shows a nontrivial real part and imaginary part (Fig. \ref{omega}). As all known, the imaginary energy part indicates a dissipated quantum state which does not have long lifetime. A positive imaginary energy ($Im E > 0$) indicates an amplifying mode and a negative imaginary energy ($Im E > 0$) ensures an attenuating mode. A stable quantum state only exists for an eigenenergy with zero imaginary part. For the small pairing energy gap, $\Delta=3$, the real energy spectrum is composed of periodical pairs of a small energy loop and a big energy loop, with a linear dispersion bridging them. Zero imaginary part of energy spectrum only exists in discrete momentum zones which correspondsto the energy loop pairs (Fig. \ref{omega} (a)). The linear modes that bridges the two loops has two fold degeneracy but are not stable, since the corresponding imaginary part forms non-zero loops. This linear mode indicate a short term motion of boson-fermion pair, whose velocity is determined by $v = i\hbar\partial_{k}E$. The biased directional motion of boson-fermion pair does not last for long time, it stops gradually over the dissipation period. Since the two neighboring linear dispersions have opposite gradient in momentum space, the boson-fermion pair moves in opposite direction in these momentum zones. For a slightly bigger energy gap, $\Delta=4$, the small loops vanished and is replaced by one energy curve. This energy curve state only has stable existence at certain momentum points, which corresponds to the zero imaginary energy points (Fig. \ref{omega} (b)). For a larger pairing energy gap, $\Delta=12$, the real energy spectrum turn into conventional energy wave, but is always accompanied by imaginary part, which only show zero points at certain wave vectors. As the pairing energy gap becomes larger than $\Delta=15$, the energy spectrum is no longer stable, the non-zero imaginary energy terms exist in the whole momentum space, which limited the lifetime of large gap states. If the fermion and boson has the same chemical potential and tunneling rate, then they share the same dispersion, $\epsilon_{k,b}=\epsilon_{k,f}$, this leads to supersymmetric state jointed with time reversal symmetry (Fig. \ref{omega} (c)). Both the real term and imaginary term of this supersymmetric spectrum are energy loops for finite gap $\Delta=15$ (Fig. \ref{omega} (c)). But the chiral linear dispersions now are parallel to the momentum axis, indicating a zero velocity of boson-fermion pair. In the extreme case of gapless phase, the imaginary energy spectrum vanished, the real energy spectrum exist as intersecting waves (Fig. \ref{omega} (d)). Each intersecting point can be approximated by Dirac cone (Fig. \ref{omega} (d)).

\begin{figure}[htbp]
\centering
\par
\begin{center}
$
\includegraphics[width=0.45\textwidth]{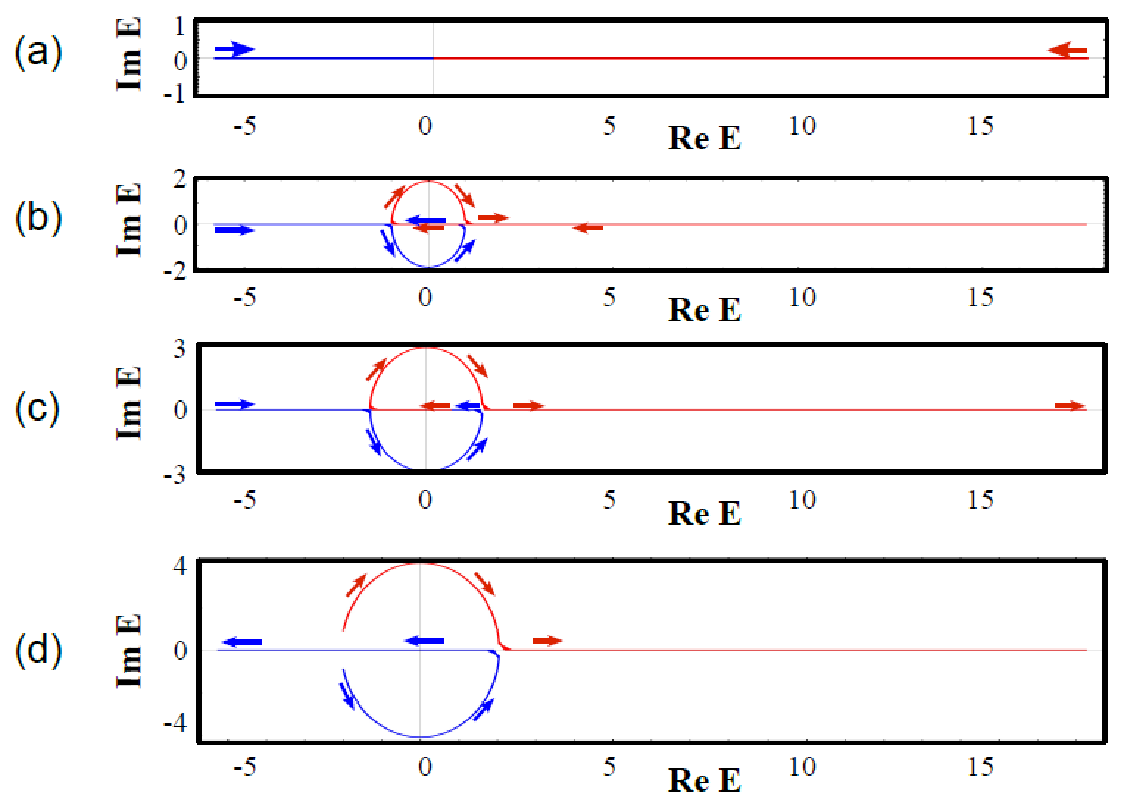}
$
\end{center}\vspace{-0.3cm}
\caption{\label{reimE} The complex energy spectrum with respect to different energy gap (a) $\Delta = 0$,(b) $\Delta = 2 $,(c) $\Delta = 3 $, (d) $\Delta = 4.1 $, with the other parameters $\mu_{f}=-1,\mu_{b}=-3$, $t_{f}=1,t_{b}=3$; Here the occupation of fermions is larger than 1/2. The red arrows indicates the energy flow with respect to an increasing momentum vector.}
\vspace{-0.2cm}
\end{figure}

Even though this boson-fermion pairing Hamiltonian is Hermitian, it generates a complex energy spectrum. The trace of the complex energy is depicted by a vector which is defined by the real and imaginary part of the complex spectrum. When the momentum vector increase from 0 to $2\pi$, this energy vector winds around the zero point in different ways under different parameters. Different gap states can be classified by a topological winding number \cite{gong},
\begin{eqnarray}\label{v}
W = \sum_{n}\frac{1}{2\pi}\int_{0}^{2\pi}\partial_{k}[\arg E_{n}(k)].
\end{eqnarray}
Here $\arg$ denotes the angle of the complex energy vector. The boson-femrion excitation has two eigen-energy levels, as depicted in Fig. (\ref{reimE}) by the red curves and blue curves. The red and blue arrows point out the energy flow direction when momentum vector $k$ increases. For the special case of gapless state, the energy starts flow from far region to the zero point Fig. (Fig. \ref{reimE} (a)), which leads to a zero winding nubmer. For finite gap $\Delta=2$, one energy branch (indicated by the blue current) swipes over $-3\pi/2$ in counterclockwise direction. The other branch winds around the origin point over an angle of $\pi$ (Fig. \ref{reimE} (b)). Thus winding number of the sum of these two windings are $ W =-1/4$, which labeled the gap state with $\Delta=2$. The gap state with $\Delta=3$ share the same winding number as that for $\Delta=2$ (Fig. \ref{reimE} (c)). When the pairing gap increase to $\Delta = 4$, the initial point of energy spectrum starts at an angle of $0.93\pi$ (the red branch) and $1.07\pi$ (the blue branch) (Fig. \ref{reimE} (d)), in this case, the winding operation does not go through a complete circle, thus the winding number for this case is an irrational number.

\begin{figure}[htbp]
\centering
\par
\begin{center}
$
\begin{array}{c@{\hspace{0.05in}}c}
%\multicolumn{1}{1}{\mbox{}} & \multicolumn{1}{1}{\mbox{}} \\
\includegraphics[width=0.23\textwidth]{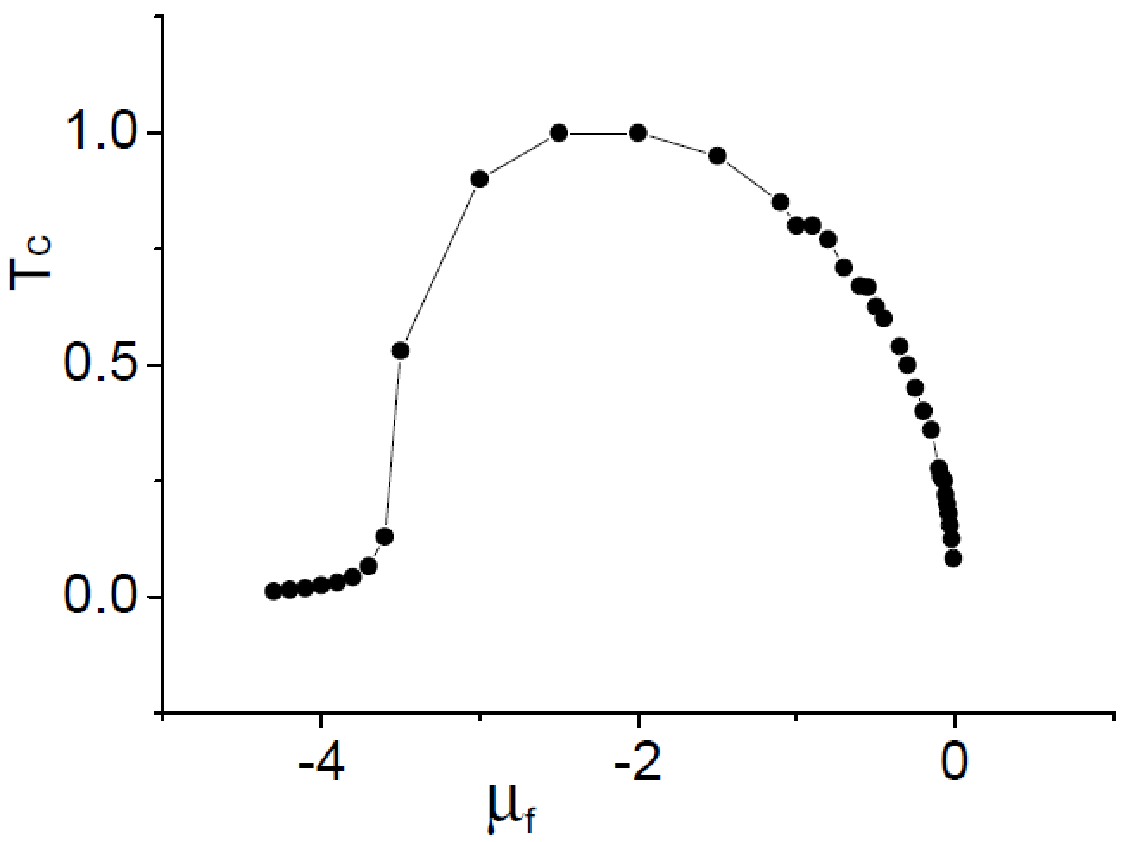}&\includegraphics[width=0.23\textwidth]{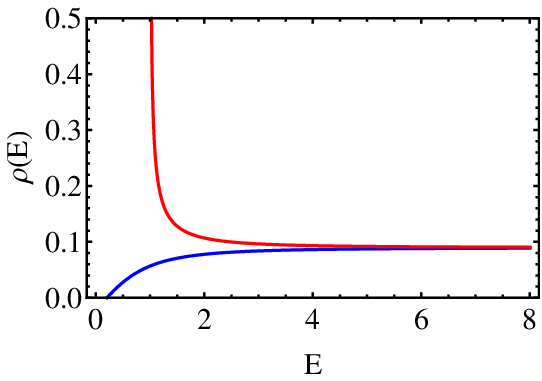}\\
\mbox{(a)} & \mbox{(b)}\\
\end{array}
$
\end{center}\vspace{-0.3cm}
\caption{\label{tc} (a) The critical temperature of gap closing point varies with negative fermionic chemical potential, but it does not exist for positive chemical potential.(b) Density of states for quasi-excitation.
$\epsilon_{-}=0.2$, $\epsilon_{+}=3$. The red curve corresponds to
$n_{_f}=1$, the blue curve corresponds to $n_{_f}=0$.}
\vspace{-0.2cm}
\end{figure}

The phase transition from a gapped phase to gapless phase is labeled by a transition from winding number of $W = -1/4$ to a zero winding number $W = 0$. The gap only closes at certain finite temperature. Green function theory at finite temperature offers the boson-fermion pairing gap that depends on the occupation of fermions and the pairing strength $V$,
\begin{eqnarray}
\Delta^{\ast}=\sum_{k}VD_{1}[\frac{1}{e^{\beta E_{1}}+1}
-\frac{1}{e^{\beta E_{2}}+1}],\;\;\;\;
\end{eqnarray}
where the coefficient $D_{1}$ is
\begin{eqnarray}
D_{1}=\frac{(2n_{f,-k}-1)V\Delta^{\ast}}{\sqrt{(\epsilon_{k,b}+\epsilon_{k,f})^2-4(2n_{f,-k}-1)N_{bf}}}.
\end{eqnarray}
The gap is closed at the critical temperature where
$\Delta=0$, then the self-consistent equation above reduces to
\begin{eqnarray}\label{criTc}
1=\sum_{k}V\frac{(2n_{f,-k}-1)}{(\epsilon_{k,b}+\epsilon_{k,f})}[\frac{1}{e^{\beta\epsilon_{k,b}}+1}
-\frac{1}{e^{-\beta\epsilon_{k,f}}+1}],\;\;\;\;
\end{eqnarray}
here $n_{f,-k}$ is the Fermi-Dirac distribution,
$n_{f,-k}=({e^{\beta\epsilon_{f}}+1})^{-1}$. The solution of this critical equation (\ref{criTc}) offers the critical temperature for gap closing, $\Delta=0$. The numerical solution of this critical equation suggested that critical transition point was found for negative chemical potential of fermions $\mu_{f}$ (Fig. \ref{tc} (a)) under following parameter setting: negative chemical potential of boson $\mu_{b} = -0.01$, positive boson tunneling rate $t_{b} = 0.01$, positive fermion tunneling rate $t_{f} = 0.01$ and a normalized interaction parameter $V = 1$. The critical temperature suggest a maximal critical temperature around $\mu_{f} = -2$ (Fig. \ref{tc} (a)). However the gap is always closed in the half-filling state of fermion. This critical transition curve shows similar configuration for other parameter zones.

One approximated expression of the critical temperature equation could be obtained for two special cases: (a) the hopping rate is much smaller than the chemical potential, $t_{b/f}<<\mu_{b/f}$. Then the ratio of fermion's kinetic energy to the boson's kinetic energy is approximately proportional to ratio of the chemical potential,$\epsilon_{k,f}/\epsilon_{k,b}=\mu_{f}/\mu_{b}$; (b) Both the chemical potential of boson and fermion is zero. Then $\epsilon_{k,f}/\epsilon_{k,b}=t_{f}/t_{b},$ The ratio of fermion's hopping rate to that of boson determines the energy ratio. The critical equation can be transformed into an integral equation including two parts. Under the cutoff at Debey frequency of fermi gas in the first part of the integral equation, i.e.,
$\hbar\epsilon_{D} >> k_{B}T$, it yields $(2n_{f,-k}(\beta\epsilon_{D}f)-1)=-\tanh[\frac{\beta\epsilon_{D}f}{2}]$.
The occupation function of fermions has only two values, $\tanh[\infty]=1,\tanh[0]=0.$ The second part of the integral equation contributes a constant that fluctuates from 0.18 to 0.55 when the
parameter pairs $[(\epsilon_{k,f}/\epsilon_{k,b}),\beta]$ varies in a wide range. For the special case of $(\epsilon_{k,f}/\epsilon_{k,b})$=2 and $\beta=3$, this constant number is $0.23$. The critical integral equation reduces to a simpler form, $[(1+(\epsilon_{k,f}/\epsilon_{k,b}))+0.23]=V\ln[\beta\epsilon_{D}].$
The critical temperature reads
\begin{eqnarray}
T_{c}=\frac{\epsilon_{D}}{k_{B}}\;\exp[{-\frac{1.23+(\epsilon_{k,f}/\epsilon_{k,b})}{V}}].
\end{eqnarray}
Thus the critical temperature increase as the pairing strength $V$ increases since the binding energy of boson fermion pair contributes the energy gap. For the special case  $t_{b/f}<<\mu_{b/f}$, the critical temperature decreases for an increasing fermion chemical potential. This approximate solution is consistent with the decreasing branch in numerical result of critical integral equation (Fig. \ref{tc}). For another special case, $\epsilon_{k,f}/\epsilon_{k,b}=t_{f}/t_{b},$ the critical temperature increases for an increasing hopping rate of boson, this means the mobility of boson could enhance the superfluidity of superconducting state. While reducing the hopping rate of fermion could also increase the critical temperature. For both of the two cases, if the energy ratio $\epsilon_{k,f}/\epsilon_{k,b}$ is negative and increasing, the critical temperature increases.

The density of state of quais-excitations is computable with the exact dispersion of quasi-excitation spectrum Eq. (\ref{e1e2}), i.e., $\rho(E)=({dE}/{d\epsilon})^{-1}$,
\begin{eqnarray}
\rho(E)=\frac{\epsilon_{-}}{\epsilon_{-}^2-\epsilon_{+}^2}+
\frac{(\epsilon_{+}^2 E)/(\epsilon_{-}^2-\epsilon_{+}^2)}{[
(\epsilon_{+}^2-\epsilon_{-}^2)\Delta^2+\epsilon_{+}^2 E^2]^{1/2}}.
\end{eqnarray}
When the occupation of fermion is $n_{_f}=1$, and ($\epsilon_{-}<0$,
$\epsilon_{+}>0$, $\epsilon_{+}^2-\epsilon_{-}^2>0$), the density of state
is divergent as a square root singularity. This is the case
of BCS model of fermion pairs. While on the other case, if the occupation of fermion is
zero, $n_{_f}=0$, the density diverges slowly in the opposite orientation of square root singularity (Fig. \ref{tc} (b)). A constant density of state shows at half-filling state of fermion. Thus most observable states are concentrated at the low energy zone.

\section{Summary}

Boson-fermion mixture of ultracold atoms in optical superlattice offers an effective simulation for hybrid quantum matters in nature. While the periodically distribution of these mixtures in artificial potential lattice can generate new physical effect. When the optical superlattice transforms from one configuration to another, it drives the mixture from the collision state to division state, or vice versa. This convective boson-fermion pairing exist as an intermediate state that can be artificially created by modulating one dimensional optical superlattice. The stable pairing states, which are indicated by energy spectrum loops in momentum space, are periodically distributed over discrete momentum vector zones. A chiral linear dispersion connects neighboring energy spectrum loops, which indicates directional motion of the pair over finite lifetime. The biased transportation of electron exist as the gapless edge state in topological insulator \cite{hasan}\cite{qi} which is also constructed in ultracold atoms in optical lattice \cite{Satija}. The momentum vector zone of this linear dispersion is controlled by the gap function which closes at a critical temperature. This critical temperature shows a maximal value at negative chemical potential of fermion. This is different from the critical temperature behavior of BCS pairing model. The Hamiltonian of the boson-fermion pairing is Hermitian. Unlike the non-Hermitian Hamiltonian system \cite{gong}, this Hermitian Hamiltonian also generates a complex spectrum. The winding number of this complex spectrum reveals topological physics in exotic pairing state and provides a different view on the intermediate state between normal state and superconductor state.

\textbf{Acknowledgment} This work is supported by "National
Natural Science Foundation of China"(Grant No. 11304062).
$*$ E-mail: tieyansi@hit.edu.cn.

\end{document}